\documentclass[aps,prd, onecolumn,showpacs,showkeys,groupedaddress,11pt]{revtex4}
\usepackage{amsmath, amsthm, amssymb, amsfonts}
\usepackage{graphicx} 
\usepackage{color}
\usepackage{wrapfig}
\usepackage{float}

\begin{document}

\title{Polymer-Fourier quantization of the scalar field revisited}
\author{Angel Garcia-Chung}
\email{angel.garcia@correo.nucleares.unam.mx} 
\author{J. David Vergara}
\email{vergara@nucleares.unam.mx} 
\affiliation{Instituto de Ciencias Nucleares, Universidad Nacional Aut\'onoma de M\'exico, A. Postal 70-543, M\'exico D.F., 04510, MEXICO}

\date{\today}

\begin{abstract}
The Polymer Quantization of the Fourier modes of the real scalar field is studied within  algebraic scheme. We replace the positive linear functional of the standard Poincar\'e invariant quantization by a singular one. This singular positive linear functional is constructed as mimicking the singular limit of the complex structure of the Poincar\'e invariant Fock quantization. The resulting symmetry group of such Polymer Quantization is the subgroup $\mbox{SDiff}(\mathbb{R}^4)$ which is a subgroup of $\mbox{Diff}(\mathbb{R}^4)$ formed by spatial volume preserving diffeomorphisms. In consequence, this yields an entirely different irreducible representation of the Canonical Commutation Relations, non-unitary equivalent to the standard Fock representation. We also compared the Poincar\'e invariant Fock vacuum with the Polymer Fourier vacuum. 
\end{abstract}

\maketitle


\section{Introduction.}
Loop Quantum Gravity (LQG) is a scheme in which the gravitational field can be kinematically quantized when treated as a gauge field \cite{LQProgram1, LQProgram2, LQProgram3}. This scheme can be extended to all matter fields appearing in the standard model such as fermions and scalar fields as well as the Yang-Mills fields \cite{LQProgram3, Thiemann1}. 

Among the results of this program can be mentioned a non-perturbative representation of the geometric observables of the gravitational field, together with a background independent implementation of the spatial diffeomorphism group.  These features provide and sustain the current interest on this program. However, there are other aspects of the program (the dynamical description, the semiclassical limit, just to mention some of them) which face serious difficulties from the mathematical and physical viewpoints. Each of these aspects is currently an area of interest in the research of Loop Quantum Gravity and they will be cured, hopefully, in the future \cite{Thiemann2}.

 These issues have motivated some authors to explore models \cite{AshtekarWillis, CorichiZapata, CorichiHamilt, Abhay, Martin} mimicking some features of the LQG scenario instead of attending the challenges of the full LQG program. The hope is that these models may help to gain understanding of some specificities of the program and thus paving the way to a full quantum description of spacetime together with matter fields. One of this models, which we refer hereinafter as Polymer-Fourier Representation (PFR) was proposed  by Viqar et al. \cite{Viqar}.

The PFR of the real (linear) scalar field is a rather different approach to the quantum description of the scalar field in flat spacetime. Its spirit departs from both, the standard Fock quantization \cite{Bogoliubov}, (for the Schr\"odinger representation see \cite{JeronimoCQ1, JeronimoCQ2}) and the Polymer-Loop quantization \cite{AshtekarLewansdoski, HannoS}, although some similarities with both representations persist. The idea of the PFR is to replace each quantum harmonic oscillator within the Fourier decomposition of the field in $\mathbb{R}^4$ flat spacetime by its polymeric analog \cite{AshtekarWillis, CorichiZapata, CorichiHamilt}. This step will change the usual results of the Fock quantization of the scalar field one of which is the modification of the Feynman propagator in such a manner that the Poincar\'e symmetry is no longer valid. Nevertheless, the Poincar\'e group and hence the standard Feynman propagator can be recovered when a parameter, introduced at hand and given by $M_{\star}$, takes high (maybe Planckian) values. This parameter is introduced with the proposition of a polymer Hamiltonian on each Fourier mode $\vec{k}$ of the form 
\begin{equation}
\widehat{H}^{poly (\vec{k})} := \frac{M_{\star}}{8} [2 - \widehat{U}^{(\vec{k})}_{2 M^{-1/2}_{\star}} -  \widehat{U}^{\dagger (\vec{k})}_{2 M^{-1/2}_{\star} }] + \frac{1}{2} \omega^2_{\vec{k}} \, \widehat{\varphi}^{2 (\vec{k})} , \label{HPoly}
\end{equation}
\noindent where $\omega_{\vec{k}} := \sqrt{\vec{k}^2 + m^2}$ and $m$ is the mass of the scalar field. With this expression the Hamiltonian of the field denoted by $\widehat{H}^{poly}$ is given by the sum 
\begin{equation}
\widehat{H}^{poly} = \sum_{\vec{k}} \widehat{H}^{poly (\vec{k})} . \label{FQH}
\end{equation}
The motivation for such Hamiltonian operator is the singular nature of the representation of the algebra of observables. In this case, the Hilbert space is constructed as the product 
\begin{equation}
{\cal H}_{poly} := \otimes_{\vec{k}} {\cal H}^{(\vec{k})}_{poly}, \label{VHilbertS}
\end{equation}
\noindent where ${\cal H}^{(\vec{k})}_{poly}$ is the Hilbert space given by $L^2(\overline{R},d\mu_{Haar})$ employed in the polymer quantization of the harmonic oscillator \cite{AshtekarWillis, CorichiZapata, CorichiHamilt}. The singular nature of the representation leads to the absence of the momentum operator of the field and hence, as well as in the mechanical case, to yield dynamics the Hamiltonian (\ref{FQH}) is proposed. 

It can be observed in (\ref{HPoly}) and (\ref{FQH}) that there is no evidence of the symmetries of this quantum field theory. Less can be said about the kinematic description where the hamiltonian plays no role. From \cite{Viqar} we learn that when the dynamics is invoked, the Poincar\'e symmetry turns out to be broken and thus can only be recovered as an effective symmetry. Anyhow, this reasoning does not tell us much about the symmetry group of the PFR. Of course, an understanding of the symmetry group of PFR can shed more precise answers in regard to for example, the Unruh effect or any other effect associated with a change of reference frame \cite{Viqar2, Kajuri}. Moreover, it is commonly accepted that the violation of the Lorentz Symmetry (LS) might be associated to quantum gravity phenomena. One of the effects of the violation of the LS is a modification of the dispersion relation for the electromagnetic field \cite{GambiniPulin} as well as in fermion fields \cite{AlfaroHugo, Alfaro}. However, most of the models consider the deviation as coming from a sort of granularity of spacetime whose origin lies on a discrete nature of the quantum geometry \cite{DanielSud} or high-derivatives terms violating CPT symmetry \cite{kostelecky}. In the present case, it is not clear whether the violation of the LS of the PFR is connected or not with a discrete nature of the spacetime. As we will see further, this symmetry-breaking comes, entirely, from the non-regular representation of the field and is not connected with any of the previous models that modify the dispersion relation. The  aforementioned replacement of the quantum harmonic oscillators turns in the non-regularity of the representation of the field and is in the core of what makes the PFR different. 

 It is worth mentioning that the polymer harmonic oscillator is a non-regular representation of the Weyl algebra of such mechanical system \cite{AshtekarWillis, CorichiZapata, CorichiHamilt}. In this case, a non-regular representation means (essentially) that some of the observables are not weakly-continuous represented on the given polymer Hilbert space. This breaks at least one of the requisites of the Stone-von Neumann theorems \cite{StoneNeumann} and therefore, we are dealing with a representation which is not unitarily equivalent to the standard Schr\"odinger quantization of the harmonic oscillator \cite{AshtekarWillis, CorichiZapata}. For systems with infinitely many degrees of freedom, the interest in non-regular representations is oriented to give a representation of the (spatial) diffeomorphism group \cite{HannoS}. As a result, this supplies a clear contrast with the standard quantization in which the symmetry group is the Poincar\'e group and the representation is a regular one \cite{JeronimoCQ1}. 

In this work, the algebraic analysis is used to derive the PFR of the real scalar field. The  Weyl algebra for the Fock quantization \cite{JeronimoCQ2, VelhinhoMTP} is considered but with a different algebraic state, i.e., a different positive linear functional $\omega_{PF}$. As a result, GNS-construction for this $\omega_{PF}$ gives a Hilbert space, named ${\cal H}_{PF}$, which is different to (\ref{VHilbertS}) and provides a non-regular and an irreducible representation of the Weyl algebra. The algebraic description is also used to study the symmetry group of the PFR of the real scalar field. 
 
We begin this paper in Section \ref{HamilAnalysis} by describing the Hamiltonian analysis of the Klein-Gordon field in a flat spacetime. The algebraic construction of the PFR is outlined in Section \ref{FDSF}. At the end of this section we study the symmetry group of the algebraic state and we proved that PFR is not Poincar\'e invariant. In Sec. \ref{Comparisson}, we study the relation of the PFR and the Standard Fock representation applying some of the techniques used in the comparison of Loop Quantization program. Finally we discuss the results in Sec. \ref{discusion}.

  
\section{Classical Scalar Field} \label{HamilAnalysis}

In what follows, we shall consider that the phase space variables of the real scalar field in a flat spacetime are decomposed into Fourier modes in order to obtain later the algebraic state associated with the PFR. The action of the symplectic 2-form together with the complex structure on the mechanical modes is derived for this purpose. The analysis of the observables of the scalar field is also provided together with the study of their change under Poincar\'e group ${\cal P}$ and the diffeomorphism group of the flat spacetime $\mbox{Diff}(\mathbb{R}^4)$. These results will be used in the next section to study the group invariance of the algebraic state. 

Consider a flat spacetime $(\mathbb{R}^4, \eta)$ where the Minkowski metric $\eta$ in cartesian coordinate is given by $\eta=\mbox{diag}(-,+,+,+)$, and it comes from the GR equations solved in the absence of matter. In this scenario, if $f \in \mbox{Diff}(\mathbb{R}^4)$, then the pullback of $\eta$, denoted by $f^{*}\eta$, is also a solution of the Einstein equations. In the $\mathbb{R}^4$ case, the spacetime has its associated group of diffeomorphisms $\mbox{Diff}(\mathbb{R}^4)$ which can admit the Poincar\'e group as an embedded subgroup.  The Poincar\'e group leaves invariant the flat metric $\eta$ and thus provides a relation between inertial observers.

 Let $\phi(X)$ be the real Klein-Gordon field, which is an scalar under the $\mbox{Diff}(\mathbb{R}^4)$ group, i.e., $\phi(X) = \phi'(X')$, where $X' = f(X)$ and $f \in \mbox{Diff}(\mathbb{R}^4)$. Notice that if $f \in {\cal P} \subset \mbox{Diff}(\mathbb{R}^4)$ the relation is still valid and gives rise to the usual transformation rule of the classical Klein-Gordon field in Minkowski spacetime \cite{Schweber}. The dynamic of $\phi(X)$ does not modifies the flat metric, i.e., $\phi(X)$ is a test field. 
 
 The classical action is given by
\begin{equation}
S[\phi]=-\frac{1}{2}\int_{\mathbb{R}^4} d^4 X \, \sqrt{- \det \eta} \left[ \partial_\mu \phi \partial^\mu \phi + m^2 \phi^2 \right], \label{CovAction}
\end{equation}
\noindent where the real and positive parameter $m$ is the  mass of the scalar field. The variational principle gives the Klein-Gordon equation 
\begin{equation}
(\Box + m^2 )\phi = 0, \label{KGEq}
\end{equation}
\noindent where $\Box:= \eta(\partial, \partial)$. Let ${\cal S}_m$ be the space of solutions of the equation (\ref{KGEq}) for a fixed value of the mass $m$. In flat spacetime the space ${\cal S}_m$ is a subset ${\cal S}_m \subset {\cal S}$ of the Schwartz space ${\cal S}$, hence each element of ${\cal S}_m$ is a rapidly decreasing smooth function. This space ${\cal S}_m$ is used to derive the phase space in which the Hamiltonian formalism takes place. The next step in this direction is to recall that $\mathbb{R}^4$ is globally hyperbolic and thus admits a foliation into spacelike Cauchy hypersurfaces \cite{Wald}. Let $e_t$ be a foliation $e_t: \mathbb{R}^3 \rightarrow \mathbb{R}^4, \vec{x} \mapsto X=e_t(\vec{x})$ such that it induces the flat space metric in the cartesian coordinates $\vec{x} \in \mathbb{R}^3$ and the lapse and shift functions are $N=1$ and $N^a = 0$ respectively. We named this type of foliations: inertial observer. Any other foliation of this type $e'_{t'}$ can be obtained by acting with the Poincar\'e group on $e_t$ yielding $e'_{t'} = (\Lambda,a) \circ e_t$ where $(\Lambda,a) \in {\cal P}$. Let us fix this inertial foliation $e_t$ to construct the phase space used for the definition of the observables of the quantum theory. The configuration field is defined as
\begin{equation}
\varphi(t; \vec{x}):= \phi(e_t) \label{FCCfield}
\end{equation}
\noindent and when considered in the action (\ref{CovAction}) for an inertial observer $e_t$, the action takes the form
\begin{equation}
S[\varphi, \dot{\varphi}] = \frac{1}{2}\int_{\mathbb{R}} dt \, \int_{\mathbb{R}^3} d^3 \vec{x} \left[ \dot{\varphi}^2 - \partial_a \varphi \partial^a \varphi - m^2 \varphi^2 \right]. \label{Taction}
\end{equation}

The canonically conjugate momentum of field $\varphi(t; \vec{x})$ and relative to the foliation $e_t$ is defined as
 \begin{equation}
 \pi(t; \vec{x}):=\dot{\varphi}(t,\vec{x})= T^\mu\partial_\mu \phi(X(e_t)). \label{MCCfield}
 \end{equation}
 Notice that (\ref{MCCfield}) is a density of weight one while the configuration field (\ref{FCCfield}) is an scalar of zeroth weight. The definition of the momentum $\pi(t; \vec{x})$ in a non-inertial foliation $\tilde{e}_{\tilde{t}}$ is given by $\tilde{\pi}(\tilde{t}; \vec{x}):=\sqrt{\tilde{q}} \; \tilde{n}^\mu(\tilde{e}) \, \nabla_\mu \phi(\tilde{e}_{\tilde{t}})$, where $\tilde{q}$ is the determinant of the spatial metric on the new foliation as well as the normal vector $\tilde{n}(\tilde{e})$. Notice that this expression can be  extended to curved spaces.
  
 Now fix the Cauchy surface $t=0$ as the initial time Cauchy surface. The initial data is the restriction $\varphi_0(\vec{x}):= \varphi(t=0;\vec{x})$ and $\pi_0(\vec{x}):= \pi(t=0;\vec{x})$ to the hypersurface $t=0$. For a given foliation $e_t$ a one-to-one correspondence between a solution $\phi(X)$ and a pair $(\varphi_0; \pi_0)$ had been obtained. Recall that $\phi$ is a Schwartz function and hence $ \varphi_0 \in {\cal S}^{(0)}(\mathbb{R}^3)$ and $ \pi_0 \in {\cal S}^{(1)}(\mathbb{R}^3)$ where ${\cal S}^{(0)}(\mathbb{R}^3)$ is the Schwartz space formed by scalars of zero-weight while ${\cal S}^{(1)}(\mathbb{R}^3)$ are the Schwartz scalars of weight one. The spaces ${\cal S}^{(0)}(\mathbb{R}^3)$ and ${\cal S}^{(1)}(\mathbb{R}^3)$ are topological vector spaces \cite{ReedSimon} and thus their topological product ${\cal S}^{(0)}(\mathbb{R}^3) \times {\cal S}^{(1)}(\mathbb{R}^3)$ is also a topological vector space. We thus omit the subindex in $\varphi_0$ and $\pi_0$ (unless it is required to avoid confusion) and define the phase space as the product space 
\begin{eqnarray}
\Gamma &:=& {\cal S}^{(0)}(\mathbb{R}^3) \times {\cal S}^{(1)}(\mathbb{R}^3) = \left\{ ( \varphi, \pi) | \, \varphi, \pi: \mathbb{R}^3 \rightarrow \mathbb{R} ; \quad  \varphi \in {\cal S}^{(0)}(\mathbb{R}^3), \quad \pi \in {\cal S}^{(1)}(\mathbb{R}^3) \right\},
\end{eqnarray}
\noindent which can be considered as a real vector space of infinite dimension. The symplectic two-form is defined as the map $\Omega: \Gamma \times \Gamma \rightarrow \mathbb{R}$ and such that  
\begin{equation}
\Omega ((\varphi_1, \pi_1), (\varphi_2 , \pi_2)):= \int_{\mathbb{R}^3} d^3x \, (\pi_1 \varphi_2 - \varphi_1 \pi_2), \label{2Form}
\end{equation}
\noindent which is a non degenerate, antisymmetric and closed map. The space $\Gamma$ together with $\Omega$ comprise the phase space $(\Gamma, \Omega) $ of the real scalar field \cite{Wald}. 

One of the aims of this work is to analyze the symmetry group of the polymer algebraic state $\omega_{PF}$ which gives rise the PFR via the GNS-construction. A detailed analysis of this issue will be given in the next section. However, to achieve this goal, it is necessary to examine the transformation rule by which a diffeomorphism transforms the observables. Of course, despite there might be several groups involved within this analysis, our attention is placed in the diffeomorphism group $\mbox{Diff}(\mathbb{R}^4)$. Therefore, now we are going to analyze the transformation of the phase space points under the action of an element of the diffeomorphism group $\mbox{Diff}(\mathbb{R}^4)$.

 Consider a different foliation $e'_{t'}(\vec{x}')$, non necessarily an inertial one with lapse, shift and spatial induced metric given by $N'$, $N'^a$ and $q'^{ab}$. Notice the new coordinates of the foliation are $(t'; \vec{x}')$, where $\vec{x}'$ is still a cartesian coordinate of some point on $\mathbb{R}^3$ and recall that there exist an element $f \in \mbox{Diff}(\mathbb{R}^4)$ such that $e' = f \circ e$ \cite{Wald}. We will simplify the notation by writing $t'=t'(t;\vec{x})$ and $\vec{x}'=\vec{x}'(t;\vec{x})$ as the action of $f $, i.e., a passive view of the diffeomorphism $f$.  In this notation the point $p \in \mathbb{R}^4$ has the coordinates $(t'; \vec{x}')$ in the foliation $e'_{t'}$ and the same point in the foliation $e_t$ is seen with coordinates $(t; \vec{x})$. Notice that a point in the surface $(t'=0; \vec{x}')$ of the foliation $e'_{t'}$ corresponds to a point $(t(t'=0; \vec{x}');\vec{x}(t'=0; \vec{x}'))$ in the foliation $e_{t}$.
 
 The scalar nature of the field $\phi$ relates the values of the field in different frames and therefore, the configuration and momentum variables of the field in both foliations are given as
\begin{eqnarray}
&&\varphi'( t'; \vec{x}') = \varphi(t(t'; \vec{x}'); \vec{x}(t'; \vec{x}')), \label{Confg} \\
&&\pi'( t'; \vec{x}') = \sqrt{q'} \; {n'}^\mu(e') \, \nabla_\mu \phi({e'}) = \frac{\sqrt{q'}}{N'} \left[ \frac{\partial}{\partial t'} - N'^a \frac{\partial}{\partial x'^a} \right] \varphi'(t'; \vec{x}'), \nonumber \\
&=& \frac{\sqrt{q'}}{N'} \left[ \frac{\partial}{\partial t'} - N'^a \frac{\partial}{\partial x'^a} \right] \varphi(t(t'; \vec{x}'); \vec{x}(t'; \vec{x}')) , \nonumber \\
&=& \frac{\sqrt{q'}}{N'} \left[ \left(  \frac{\partial t}{\partial t'} - N'^a \frac{\partial t}{\partial x'^a}  \right) \pi( t(t'; \vec{x}'); \vec{x}(t'; \vec{x}')) + \left(  \frac{\partial \vec{x}}{\partial t'} - N'^a \frac{\partial \vec{x}}{\partial x'^a}  \right)\cdot (\vec{\nabla} \varphi)(t(t'; \vec{x}'); \vec{x}(t'; \vec{x}')) \right], \label{MCano}
\end{eqnarray}
\noindent and notice that in the third line we inserted (\ref{Confg}). It can be seen from (\ref{Confg}) and (\ref{MCano}) that in general, initial data $(\varphi'_0, \pi'_0)$ corresponds to `evolved' data in the time $t(t'=0; \vec{x}')$ within foliation $e_{t}$, i.e., it is not necessarily initial data in $e_{t}$. 

Now we want to express the transformed data given in (\ref{Confg})-(\ref{MCano}) in terms of the initial data (\ref{FCCfield})-(\ref{MCCfield}). To do so, recall that the one particle Hilbert space, used to derive the Fock space of the free Klein-Gordon field, is built up with solutions of the Klein-Gordon equation (\ref{KGEq}). This feature allows us to use the Hamilton equations as well. 

To obtain the Hamilton equations we first define the hamiltonian density for the system which is given by 
\begin{equation}
 H[\varphi, \pi] = \frac{1}{2} \pi^2 + \frac{1}{2} \varphi \left( - \Delta + m^2 \right) \varphi,
\end{equation}
\noindent and is defined via the Legendre transformation associated to the momentum on the action (\ref{Taction}). It gives the Hamilton equations of the form
\begin{eqnarray}
\dot{\varphi} &=& \pi , \\
\dot{\pi} &=& - \widehat{\Theta} \, \varphi,
\end{eqnarray}
\noindent where $\widehat{\Theta} := - \Delta + m^2$ and $\Delta := \partial_a \partial^a$. The solutions to these equations are of the form
\begin{eqnarray}
&&\varphi(t; \vec{x}) =  \widehat{A}[t;\vec{x}] \, \varphi(0; \vec{x}) + \widehat{B}[t;\vec{x}] \, \pi(0; \vec{x}) , \label{fieldt} \\
&& \pi(t; \vec{x}) = \widehat{C}[t;\vec{x}] \, \varphi(0; \vec{x}) + \widehat{A}[t;\vec{x}]\, \pi(0; \vec{x}), \label{moment}
\end{eqnarray}
 \noindent where the operators $\widehat{A}$, $\widehat{B}$ and $\widehat{C}$ are defined as
 \begin{eqnarray}
\widehat{A}[t;\vec{x}] &:=& \cos\left[ t \sqrt{\widehat{\Theta}}\right] , \label{AOperator} \\ 
\widehat{B}[t;\vec{x}]&:=& \widehat{\Theta}^{-\frac{1}{2}} \sin\left[ t \sqrt{\widehat{\Theta}}\right], \label{BOperator} \\
\widehat{C}[t;\vec{x}]&:=& - \widehat{\Theta}^{\frac{1}{2} }\sin\left[ t \sqrt{\widehat{\Theta}}\right]  . \label{COperator}
\end{eqnarray}

By inserting (\ref{fieldt}) and (\ref{moment}) in (\ref{Confg}) and (\ref{MCano}), it is possible to express the initial data of the foliation $e'_{t'}$ in terms of the initial data of the foliation $e_t$ in the form
\begin{eqnarray}
\varphi'( 0; \vec{x}') &=& \widehat{A}[t(0; \vec{x}'); \vec{x}(0; \vec{x}')] \, \varphi_0 + \widehat{B}[t(0; \vec{x}'); \vec{x}(0; \vec{x}')]\, \pi_0 , \label{FieldDif} \\
\pi'( 0; \vec{x}') &=& \widehat{D}[t(0; \vec{x}'); \vec{x}(0; \vec{x}')] \, \varphi_0 + \widehat{E}[t(0; \vec{x}'); \vec{x}(0; \vec{x}')] \, \pi_0 . \label{MomenDif}
\end{eqnarray}
The operators $\widehat{D}$ and $\widehat{E}$ are defined as
\begin{eqnarray}
\widehat{D} &:=& \frac{\sqrt{q'}}{N'} \left[  \left( \frac{\partial t}{\partial t'} - N'^a \frac{\partial t}{\partial x'^a} \right) \widehat{C} +  \left( \frac{\partial \vec{x}}{\partial t'} - N'^a \frac{\partial \vec{x}}{\partial x'^a} \right)\cdot \nabla (\widehat{A} \cdot) \right] , \label{DOperator} \\
\widehat{E}&:=&  \frac{\sqrt{q'}}{N'} \left[  \left( \frac{\partial t}{\partial t'} - N'^a \frac{\partial t}{\partial x'^a} \right) \widehat{A} +  \left( \frac{\partial \vec{x}}{\partial t'} - N'^a \frac{\partial \vec{x}}{\partial x'^a} \right)\cdot \nabla (\widehat{B} \cdot) \right], \label{EOperator}
\end{eqnarray}
\noindent where the notation $\nabla (\widehat{O} \cdot)$ indicates that $\nabla (\widehat{O} \cdot) P(x) := \nabla (\widehat{O} P)$.  

Summarizing, a diffeomorphism $f$, acting on the inertial foliation $e_t$ gives rise the new foliation $e'_{t'}$ and `moves' the initial data $(\varphi_0, \pi_0)$ to the initial data $(\varphi'_0; \pi'_0)$ given by (\ref{FieldDif}) and (\ref{MomenDif}). In others words, a diffeomorphism $f$ induces a transformation $T_f$ on the phase space which can be written succinctly as
\begin{equation}
T_f : \Gamma \rightarrow \Gamma; (\varphi, \pi) \mapsto (\varphi', \pi'):= (\widehat{A} \varphi + \widehat{B} \pi, \widehat{D}\varphi + \widehat{E} \pi), \label{TIndF}
\end{equation}
\noindent where the operator coefficients $\widehat{A}$, $\widehat{B}$, $\widehat{D}$ and $\widehat{E}$ contain all the information of the diffeomorphism $f$ and are given in (\ref{AOperator}-\ref{BOperator}) and (\ref{DOperator}- \ref{EOperator}).

Let us now define the observables of the real scalar field which will be used in the quantum description. As is already known \cite{JeronimoCQ1, JeronimoCQ2}, to avoid some ambiguities in the quantization program (such as ordering problems) with a general selection of observables, we pick out only linear observables. Therefore we consider as observables of the real scalar field, real linear (and continuous) functions on the phase space $\Gamma$. The selection of observables is directly related with the labels employed for the Weyl generators as we will see in section \ref{FDSF}. A quantum theory based on different sets of Weyl generators gives rise, at least intuitively, to different quantum theories of the same classical system. 
  
 Consider the observables given by
\begin{eqnarray}
 F_\lambda[(\varphi, \pi)]&:=& \Omega ((g_{\varphi},g_{\pi}), (\varphi , \pi))  = \int_{\mathbb{R}^3} d^3 \vec{x} (g_{\pi} \varphi - g_{\varphi} \pi), \; \; \lambda:= (g_{\varphi},g_{\pi}) \in \Gamma, \label{FTObv}
\end{eqnarray}
\noindent where the label $\lambda$ is an element of the phase space and thus is a pair of Schwartz functions. It can be checked directly that the space comprised by these observables, denoted by ${\cal F}$, is a real vector space. The symplectic $2-$form $\Omega$ given in (\ref{2Form}) can be extended by linearity to ${\cal F}$, thus providing ${\cal F}$ with a real symplectic vector space nature. The induced symplectic form is given by
\begin{eqnarray}
\Omega(F_\lambda , F_\mu) &=& \Omega(F_{(g_{\varphi}, g_{\pi})} , F_{(\tilde{g}_{\varphi},\tilde{g}_{\pi})}) =  \int_{\mathbb{R}^3} d^3 \vec{x} \, ( g_{\pi} \, \tilde{g}_{\varphi} - g_{\varphi} \, \tilde{g}_{\pi} ).
\end{eqnarray}
These observables are commonly used in the standard quantization procedure given for example in \cite{JeronimoCQ2, JeronimoCQ1, Wald, VelhinhoMTP, Jackiw}. We will see in section \ref{Comparisson} that the sympletic space ${ \cal F}$ gives rise to the standard Weyl algebra of the Quantum Klein Gordon field.

Recall now that a diffeomorphism $f$ induces the transformation $T_f$ given in (\ref{TIndF}) on the phase space $\Gamma$. This transformation induces a new transformation $A_{T_f}$ on the observables $F_{\lambda}$ which we define as
\begin{eqnarray}
A_{T_f} : {\cal F} \rightarrow {\cal F}; F_{\lambda}=F_{(g_\varphi , g_\pi )} \mapsto \tilde{F}_{\lambda} := F_{T_f (\lambda)} = F_{T_f(g_\varphi , g_\pi )}. \label{TSObs}
\end{eqnarray}  
 \noindent Expression (\ref{TSObs}) will be used in the analysis of the symmetry group of the PFR. Recall that the definition of the observables $F_\lambda$ was based on the space of solutions,  which in turn, was employed to define the phase space. 

Let us now move to the Fourier decomposition of the phase space $\Gamma$. Consider a torus kind topology for the base space $\mathbb{R}^3$ and a box of finite but fixed volume $V$ on it \cite{Wald, JeronimoCQ1}. The scalar field $\varphi$ and its momentum $\pi$ when restricted to this box can be decomposed in Fourier modes which taking the form
\begin{eqnarray}
\varphi(t; \vec{x}) &=&\sum_{\vec{k}}  \frac{\left[ (1+i) q_{\vec{k}}(t) + (1-i) q_{-\vec{k}}(t) \right]}{2\sqrt{V}}  e^{i \vec{k} \cdot \vec{x}} , \label{scalarF}\\
\pi(t; \vec{x}) &=& \sum_{\vec{k}} \frac{ \left[ (1+i) p_{\vec{k}}(t) + (1-i) p_{-\vec{k}}(t) \right] }{2\sqrt{V}}  e^{i \vec{k} \cdot \vec{x}}. \label{momscalar}
\end{eqnarray}   
This particular description implies that the Hamiltonian of the field is described as the sum of the Hamiltonians of harmonic oscillator for each Fourier mode. Naturally, the variables $q_{\vec{k}}$ and $p_{\vec{k}}$ satisfies the CCR algebra of the harmonic oscillator
\begin{equation}
\{ q_{\vec{k}} , q_{\vec{k}'}\} = \{ p_{\vec{k}} , p_{\vec{k}'}\} = 0, \hspace{1cm} \{ q_{\vec{k}} , p_{\vec{k}'}\} = \delta_{\vec{k}, \vec{k}'}.
\end{equation}

 It is important to mention that adequate properties to transform the Dirac delta into a Kronecker delta function and the converse are necessary. In our case they take the form
\begin{equation} \label{Prop}
\int d^3 x \, \frac{e^{i (\vec{k} + \vec{k}') \cdot \vec{x}}}{V}   = \delta_{\vec{k} + \vec{k}',\vec{0}}, \hspace{0.5cm}  \sum_{\vec{k}} \frac{e^{i \vec{k} \cdot (\vec{x} - \vec{y})}}{V} = \delta^{(3)}(\vec{x} - \vec{y}).
\end{equation}

These expressions allow us to write the mechanical variables in terms of the field as 
\begin{eqnarray} \label{DecMode}
q_{\vec{k}} &=& \int \frac{d^3 x}{\sqrt{V}}\, \varphi(\vec{x}) \left[ \cos(\vec{k} \cdot \vec{x}) - \sin(\vec{k} \cdot \vec{x}) \right] , \\
p_{\vec{k}} &=& \int \frac{d^3 x}{\sqrt{V}}\, \pi(\vec{x}) \left[ \cos(\vec{k} \cdot \vec{x}) - \sin(\vec{k} \cdot \vec{x}) \right].
\end{eqnarray}

As can be seen, the expressions given in (\ref{scalarF}) - (\ref{momscalar}) characterizes the field and its momentum by the coordinates and mechanical momentum variables, i.e., every field $\varphi(\vec{x})$ is uniquely determined by the set of variables $ \{ q_{\vec{k}} \} $, and the momentum field $ \pi(\vec{x}) $ by the corresponding set of variables $ \{ p_{\vec{k}} \} $ where the variables $\{ q_{\vec{k}}\}$ and $\{ p_{\vec{k}}\}$ are the coordinates of the field $\varphi$ and $\pi$ respectively in Fourier basis $\{ \frac{e^{i \vec{k} \cdot \vec{x}}}{V} \}$. This is confirmed in the expressions given in (\ref{DecMode}) showing the relations between the coordinates $(q_{\vec{k}},p_{\vec{k}})$ and any particular value of the field and its momentum. With this description every point $(\varphi(\vec{x}) , \pi(\vec{x}))$ of the phase space corresponds to the set $\{ (q_{\vec{k}} , p_{\vec{k}})\}$.

The next step in the construction of Fock representation for the scalar field theory is to consider the complex structure $J$. Before to do so, let us summarize the idea behind (for details see \cite{AbhayMagnon, JeronimoCQ1, Wald}). 

The Fock Hilbert space ${\cal F}_{Fock}$ is built up via a symmetrize infinite product ${\cal F}_{Fock} := \otimes^{(S)}_n {\cal H}^n$ of what is called the `one-particle Hilbert space' ${\cal H}$. This one-particle Hilbert space on the other hand is constructed from the complexification of the phase space $\Gamma$. It is necessary to split the complexified phase space $\Gamma^{\mathbb{C}}$ into positive and negative frequencies. The one formed with positive frequencies is used to build up the one-particle Hilbert space. 

 Naturally, there are infinitely many ways of splitting the complexified phase space into positive and negative frequencies. Each of them can be obtained via a complex structure $J$. The complex structure is a map $J: \Gamma \rightarrow \Gamma, (\varphi, \pi) \mapsto (\varphi_J , \pi_J) := J(\varphi , \pi)$ preserving the symplectic form $\Omega$ and such that $J^2 = -1$. In the context of flat spacetime where the Poincar\'e group is available, there is a unique complex structure $J^{(P)}$ which is invariant under the Poincar\'e group action  \cite{AbhayMagnon, JeronimoCQ1} given by 
\begin{equation}
J^{(P)} (\varphi, \pi) = (- (-\Delta + m^2)^{-1/2} \pi , (-\Delta + m^2)^{1/2} \varphi).
\end{equation}

 Remarkably, any other complex structure $J$ in flat spacetime compatible with the Poincar\'e group will give rise to a Fock type representation which is unitarily equivalent to the previous one \cite{JeronimoCQ1}. This result is closely related to the fact that the Poincar\'e group does not change the values of the lapse, the shift and the spatial metric as we already stated \cite{AbhayMagnon}. Thus $J^{(P)}$ is the unique complex structure available for this system which gives rise to the (Poincar\'e invariant) Fock representation and will be used to derive the algebraic state corresponding to the PF quantization.

 We now insert (\ref{scalarF}) - (\ref{momscalar}) in the corresponding expression for the symplectic form of the field $\Omega$ and using the properties (\ref{Prop}) we get the following expression for the symplectic form $\Omega$ in terms of the symplectic form $\Omega^{(\vec{k})}$ of the mechanical $\vec{k}$-modes 
\begin{eqnarray}
&& \Omega((\varphi, \pi), (\tilde{\varphi}, \tilde{\pi})) = \sum_{\vec{k}} \Omega^{(\vec{k})}((q_{\vec{k}}, p_{\vec{k}}),(\tilde{q}_{\vec{k}}, \tilde{p}_{\vec{k}})), \qquad \Omega^{(\vec{k})}((q_{\vec{k}}, p_{\vec{k}}),(\tilde{q}_{\vec{k}}, \tilde{p}_{\vec{k}})) := p_{\vec{k}} \tilde{q}_{\vec{k}}- q_{\vec{k}}\tilde{p}_{\vec{k}}.
\end{eqnarray}

This result is important because allow us to describe $\Omega$ in terms of a linear relation of the $\Omega^{(\vec{k})}$'s which is a consistent result if we insist to think the free scalar field and its momentum as a tower of infinite harmonic oscillators on each point $\vec{x} \in \mathbb{R}^3$.

As well as the symplectic form can be decomposed into Fourier modes, the complex structure also admits such decomposition. $J^{(P)}$ can also be described in terms of the complex structure of each harmonic oscillator denoted by $J^{(\vec{k})}$ as follows
\begin{eqnarray} 
&& J^{(P)} \left(  \varphi,  \pi  \right) = \sum_{\vec{k}}  \frac{J^{(\vec{k})} e^{i \vec{k} \cdot \vec{x}}}{2\sqrt{V}}  \left( \begin{array}{c} (1+i) q_{\vec{k}} + (1-i) q_{-\vec{k}} \\ (1+i) p_{\vec{k}} + (1-i) p_{-\vec{k}} \end{array} \right) , \label{Jmode} \\
 && J^{(\vec{k})} := \left( \begin{array}{cc} 0 & \omega_{\vec{k}} \\ - \omega^{-1}_{\vec{k}} & 0 \end{array}\right) , \label{JK}
\end{eqnarray}
\noindent and where the parameter $\omega_{\vec{k}} := \sqrt{\vec{k}^2 + m^2}$ is the frequency of the mode $\vec{k}$. Recall that the $J^{(\vec{k})}$ map acts on the $\vec{k}$-mode variables as
\begin{equation}
\left( q_{\vec{k}}, p_{\vec{k}} \right) \mapsto \left( q^J_{\vec{k}}, p^J_{\vec{k}} \right) := J^{(\vec{k})}\left( q_{\vec{k}}, p_{\vec{k}} \right) = \left( \omega_{\vec{k}} p_{\vec{k}} , - \frac{1}{\omega_{\vec{k}} } q_{\vec{k}}\right).
\end{equation}
  
  Finally, it will be important the Fourier decomposition of observables $F_\lambda$. It is easy to check that if we define
  \begin{eqnarray}
  g(\vec{x}) &=&\sum_{\vec{k}}  \frac{\left[ (1+i) g_{\vec{k}} + (1-i) g_{-\vec{k}} \right]}{2\sqrt{V}}  e^{i \vec{k} \cdot \vec{x}} , \\
f(\vec{x}) &=& \sum_{\vec{k}} \frac{ \left[ (1+i) f_{\vec{k}} + (1-i) f_{-\vec{k}} \right] }{2\sqrt{V}}  e^{i \vec{k} \cdot \vec{x}}, 
  \end{eqnarray}
  \noindent then the standard observables take the form
  \begin{equation}
  F_{\lambda} = \sum_{\vec{k}} \left( q_{\vec{k}} f_{\vec{k}} - p_{\vec{k}}  g_{\vec{k}}  \right) =: F_{ (\{ g_{\vec{k}} \}, \{ f_{\vec{k}} \} )}.
  \end{equation}

In the limit of $V \rightarrow \mathbb{R}^3$ the coefficients $g_{\vec{k}}$ and $f_{\vec{k}}$ tend to be the Fourier transform of the Schwartz functions $g$ and $f$. Of course, in this limit the functions $g({\vec{k}})$ and $f({\vec{k}})$ are also Schwartz functions. This result will play an important role in the analysis of the irreducibility of the PF representation given by Viqar et al.


\section{Algebraic analysis of the Polymer Fourier representation}\label{FDSF}

To study the representations of the CCR, the Weyl algebra is the appropriate arena. By using the GNS-construction for the Weyl algebra, any Fock type representation based on the CCR is obtained \cite{Wald, Bratelli, Haag, Corichi2}. First let us review the GNS-construction which has as principal ingredients a $C^*$ unital algebra ${\cal A}$ of observables and a given state $\omega$. States are positive linear functionals $\omega: {\cal A} \rightarrow \mathbb{C}$ satisfying the conditions
 \begin{equation}
 \omega(A^* A) \geq 0 \quad \forall \hspace{0.1cm} A \in {\cal A}, \qquad \omega(\tilde{1})=1,
 \end{equation}
 \noindent where $\tilde{1}$ is the unit element in the algebra ${\cal A}$. With these two ingredients the GNS-construction gives the triplet $({\cal H}_{\omega}, \pi_\omega, \Omega_\omega)$ where ${\cal H}_{\omega}$ is a Hilbert space, $ \pi_\omega$ a representation of the $C^*-$algebra ${\cal A}$ and $\Omega_\omega$ is a cyclic vector in ${\cal H}_{\omega}$. 
 
 The Hilbert space is obtained as the completion of the pre-Hilbert space formed as the quotient ${\cal A}/{\cal J}$. The space ${\cal J}$ is the Gel'fand ideal consisting of all elements $ a \in {\cal A}$ such that $\omega(a^* a) = 0$. The representation $\pi_\omega : {\cal A} \rightarrow {\cal L}({\cal H}_\omega)$ is of the form
 \begin{equation}
 \pi_\omega(A)[B] := [A B], 
 \end{equation}
 \noindent where $[A], [B] \in {\cal H}_\omega$ and $A \in {\cal A}$. The action of ${\cal A}$ on the vector $\Omega_\omega := [\tilde{1}]$ defines a dense subset in ${\cal H}_\omega$ whence $\Omega_\omega$ is a cyclic vector. The inner product of two vectors is given by
 \begin{equation}
\langle [A] | [B] \rangle := \omega(A^* B) .
 \end{equation} 
 
 In the case of the real scalar field the $C^*$ unital algebra employed is the Weyl algebra constructed by following Slawny theorem \cite{Bratelli, slawny}. Slawny's Theorem establish the minimal Weyl algebra associated with a real symplectic vector space. It is at this point when we bring into consideration the symplectic space ${\cal F}$. The standard quantization of the scalar field considers the space given in (\ref{FTObv}). In this case, the generators of the Weyl algebra are denoted by $ \widehat{W}(\lambda) $. They are employed to define the product and the involution operation of the algebra as
\begin{equation}
\widehat{W}(\lambda) \widehat{W}(\lambda') = e^{\frac{i}{2} \Omega(\lambda,\lambda')} \widehat{W}(\lambda + \lambda'), \quad \widehat{W}(\lambda)^* = \widehat{W}(-\lambda), \label{WProd}
\end{equation}
\noindent where the presence of the symplectic $2-$form $\Omega$ in the exponential (as result of \cite{slawny}), is directly related to considering the standard observables (\ref{FTObv}).  This algebra can be promoted to a $C^*$-algebra known as Weyl algebra of the real scalar field and is denoted by ${\cal W}$.

 The algebraic state related with the Poincar\'e invariant Fock representation, denoted by $\omega_{J^{(P)}}$ is of the form
\begin{equation} 
\omega_{J^{(P)}}(\widehat{W}(\lambda)) = e^{-\frac{1}{4} \Omega(\lambda, J^{(P)}\lambda)} , \label{Fstate} 
\end{equation}
\noindent where $J^{(P)}$ was defined in the previous section. The GNS-construction of this algebra based on this positive linear functional gives rise to the standard Hilbert space $L^2({\cal S}^{'(1)}, d\mu_G)$ where $\mu_G$ is a gaussian measure that emerges as result of considering the configuration polarization after applying Bochner-Minlos Theorem \cite{JeronimoCQ1, JeronimoCQ2}. 

Now, recall that PF representation is constructed as a replacement of the quantum harmonic oscillator of the Fourier decomposition of the scalar field by its polymer analog. Our proposal is basically to consider that such replacement is equivalent to replace the algebraic state $\omega_{J^{(P)}}$ by a singular one. To construct such singular state we will use the idea of the polymer construction given in \cite{CorichiZapata}. 

In \cite{CorichiZapata}, Corichi et al deduced the polymer representations of the harmonic oscillator from the standard Schr\"odinger representation via a singular limit of the parameter appearing in the complex structure. Such parameter, denoted by $d$, admits two singular limits named `A-type' which is of the form $1/d \rightarrow 0$ and the `B-type'  which is given by $d \rightarrow 0$. In the first case we consider the $p-$polarization in which the polymer Hilbert space is of the form ${\cal H}_p := L^2(\overline{\mathbb{R}}, d\mu_{Bohr})$ while in the second case the $q-$polarization gives a Hilbert space of the form ${\cal H}_q := L^2(\mathbb{R}_d, d\mu_c)$. These polymer Hilbert spaces may emerge from the GNS construction of the mechanical Weyl algebra with generators $\widehat{W}(x,p)$ and singular algebraic states $\omega_p$ and $\omega_x$. These algebraic states are constructed using the standard algebraic state $\omega_d$ of the harmonic oscillator
\begin{equation}
\omega_d(\widehat{W}(x, p)) = e^{-\frac{1}{4} \Omega((x, p); J(x, p))} = e^{-\frac{1}{4} \left( \frac{1}{d^2} x^2 + \frac{d^2}{\hbar^2} p^2 \right)}, \label{MechState}
\end{equation}
\noindent and taking the singular limits after which the algebraic states take the form
\begin{eqnarray}
\omega_p(\widehat{W}(x,p)) &:=& \lim_{1/d \rightarrow 0} \omega_d(\widehat{W}(x, p))  = \delta_{p,0},  \label{SingMechState}\\
\omega_x(\widehat{W}(x,p)) &:=& \lim_{d \rightarrow 0} \omega_d(\widehat{W}(x, p))  = \delta_{x,0}.
\end{eqnarray}
As we already mentioned, the first state gives rise to the polymer representation in the $p-$polarization while the second to the polymer  $x-$polarization. Both yield non-regular representation of the CCR in the Hilbert spaces given by ${\cal H}_p$ where $\overline{\mathbb{R}}$ is the Bohr-compactification of the real line \cite{Velhino} and ${\cal H}_q $ where $\mathbb{R}_d$ is the real line equipped with the discrete topology. 

Recall that the labels of the standard observables, $\lambda$, are elements of the space $\Gamma$, and therefore, admit a Fourier decomposition similar to that of the scalar field (\ref{scalarF} - \ref{momscalar}). Let us proceed to decompose functions $g_{\varphi}$ and $g_\pi$ into Fourier modes and insert them together with (\ref{Jmode}) in the scalar product given in (\ref{Fstate}) to obtain the expression
\begin{eqnarray} 
\omega_{J^{(P)}}(\widehat{W}(\{q_{\vec{k}} , p_{\vec{k}} \})) &=& e^{-\frac{1}{4} \sum_{\vec{k}} \Omega^{(\vec{k})}((q_{\vec{k}}, p_{\vec{k}}),J_{\vec{k}}(\tilde{q}_{\vec{k}}, \tilde{p}_{\vec{k}}))} = \prod_{\vec{k}}{e^{-\frac{1}{4} \left( \frac{1}{\omega_{\vec{k}}} q^2_{\vec{k}} + \omega_{\vec{k}} p^2_{\vec{k}} \right)}}. \label{FstateMod}
\end{eqnarray}
Note that high energy modes $\vec{k}$ will approximate to the regime in which the mechanical state tends to be singular in the sense given in \cite{CorichiZapata}. Naturally, high energy modes are closed to the regime in which quantum effects of the geometry become important. If the field is massless then low values of the energy modes give rise also to singular values in the mechanical algebraic states, again in the sense of \cite{CorichiZapata}. In the first case this issue is related to the problem of ultraviolet divergences and in the second is related with the infrared ones, both problems are out of the scope of this work.

Formula (\ref{FstateMod}) relates the algebraic state Poincar\'e invariant $\omega_{J^{(P)}}$ in terms of the mechanical states related to the Fourier modes of the scalar field. If we consider \cite{CorichiZapata} then it is clear that the `$d^2$' parameter employed there is analog to the frequency $\omega_{\vec{k}}$ presented here. 

In this scenario, we cannot perform such limit due to the fixed value that takes the $\omega_{\vec{k}}$ on each Fourier mode. Therefore, the spirit in this construction is to notice the analog of what would be the algebraic state if such limit would be done. After noticing the possible outcome, we just replace the Poincar\'e invariant state by the inferred algebraic state. In our case this yields
\begin{eqnarray}
\omega_{\varphi-pol}(\widehat{W}(\{q_{\vec{k}} , p_{\vec{k}} \})) &=&  \prod_{\vec{k}}{\delta_{q_{\vec{k}},0}} = \delta_{\{ q_{\vec{k}} \},0 } = \delta_{\varphi,0 },  \label{CoordPol}\\
\omega_{\pi-pol}(\widehat{W}(\{q_{\vec{k}} , p_{\vec{k}} \})) &=& \prod_{\vec{k}}{\delta_{p_{\vec{k}},0}} = \delta_{\{p_{\vec{k}} \},0} = \delta_{\pi,0}. \label{MomentumPol}
\end{eqnarray}

The previous analysis can now be stated as follows: `the replacement of the harmonic oscillators of the Fourier decomposition of the scalar field by polymer harmonic oscillators within each $\vec{k}-$mode is the same {\sf operation} as to replace the Poincar\'e invariant positive linear functional $\omega_{J^{(P)}}$ by the singular algebraic states (\ref{CoordPol}) or (\ref{MomentumPol}). The first one is equivalent to replace the harmonic oscillator for its polymer version but in the $q -$polarization while the second is equivalent to consider the polymer harmonic oscillators in the $p -$polarization.'

As can be seen, this representation cannot be used to derived the CCR from the Weyl algebra because is not a weakly continuous function on $\{ q_{\vec{k}} \} $ for (\ref{CoordPol}) or in $\{ p_{\vec{k}} \}$ in the case of (\ref{MomentumPol}). In this very specific sense, this is a non regular (a singular) representation of the Weyl algebra and we named Polymer-Fourier representation of the scalar field. We will see later on that this gives, formaly, the Polymer Quantum Field Theory given in \cite{Viqar}.

To derive the Hilbert space that emerges from the GNS-construction employing the states defined in (\ref{CoordPol} - \ref{MomentumPol}) we consider \cite{CorichiZapata}.  Let us begin with (\ref{CoordPol}) in which the state induces the representation of the unitary groups formed with elements $\widehat{U}(\pi):= \widehat{W}(0,\pi)$ and $\widehat{V}(\varphi):= \widehat{W}(\varphi,0)$ as
\begin{equation}
\omega_{\varphi - pol}(\widehat{U}(\pi)) = 1, \quad \omega_{\varphi - pol}(\widehat{V}(\varphi)) = \delta_{\varphi,0},
\end{equation}

\noindent which correspond to the `$\varphi$-polarization' in analogy to the mechanical case described in \cite{CorichiZapata}. For the second case we obtain the `$\pi$-polarization' given by
\begin{equation} 
\omega_{\pi - pol}(\widehat{U}(\pi)) = \delta_{\pi,0}, \quad \omega_{\pi - pol}(\widehat{V}(\varphi)) = 1. \label{ASPiPol}
\end{equation}

We can build up the Hilbert space using this result as the infinite product of polymer Hilbert spaces one for each Fourier mode as
\begin{eqnarray}
{\cal H}_{\varphi-pol} &=& \prod_{\vec{k}} {\cal H}^{(\vec{k})}_{q-poly} , \quad {\cal H}^{(\vec{k})}_{q-poly} = L^2(\mathbb{R}_d , d\mu_c) ,\label{HQPoly} \\
{\cal H}_{\pi-pol} &=& \prod_{\vec{k}} {\cal H}^{(\vec{k})}_{p-poly} , \quad {\cal H}^{(\vec{k})}_{p-poly} = L^2(\overline{\mathbb{R}} , d\mu_{Bohr}) . \label{HPPoly}
\end{eqnarray}

The Hilbert space given in (\ref{HPPoly}) is the one obtained in \cite{Viqar} at kinematical level and was already given in (\ref{VHilbertS}). In the standard Fock quantization it is possible to use a Hilbert space built up by the infinite product of the Hilbert spaces of each Fourier mode. However, the resulting representation turns out to be reducible \cite{Wald}. In the present context of Polymer Fourier representation the same happens. This means that ${\cal H}_{\pi - pol}$ is too large to yield an irreducible representation of the Weyl algebra ${\cal W}$. Let us see this in detail but first recall that if the representation is irreducible then any vector can be considered as a cyclic vector. If a vector is said to be cyclic, the action of the algebra on this vector yields a dense subspace in the Hilbert space.

Let us first recall that an element of the plane wave basis in ${\cal H}^{(\vec{k})}_{p-poly}$ corresponds to the plane wave with un-contable label $\lambda^{(\vec{k})}$ given by $N_{\lambda^{(\vec{k})}} = e^{i \lambda^{(\vec{k})} p^{(\vec{k})}}$. An arbitrary element on ${\cal H}^{(\vec{k})}_{p-poly}$ is then of the form $\Psi^{(\vec{k})} = \sum_{\lambda_j^{(\vec{k})}} \Psi_{\lambda_j^{(\vec{k})}} N_{\lambda_j^{(\vec{k})}}$ where $\Psi_{\lambda_j^{(\vec{k})}}$ is no zero on a contable set of points $\lambda^{(\vec{k})}_j$. Therefore, an arbitrary element on ${\cal H}_{\pi-poly}$ is of the form
\begin{equation}
\Psi := \prod_{\vec{k}} \Psi^{(\vec{k})},
\end{equation}
\noindent where the label $\vec{k}$ runs on continuous values ( ${\cal H}_{p-poly}$ is a non-separable Hilbert space: continuous tensor product of non-separable Hilbert spaces ${\cal H}^{(\vec{k})}_{poly}$). Notice that a plane wave basis on ${\cal H}_{p-poly}$ takes the form 
\begin{equation}
\cdots \otimes N_{  \lambda^{    (\vec{k}_1) }_{a_{\vec{k}_1}}  } \otimes N_{  \lambda^{    (\vec{k}_2) }_{a_{\vec{k}_2}}  } \otimes \cdots = e^{i \sum_{\vec{k}} \lambda^{ (\vec{k}) }_{a_{\vec{k}}} p^{(\vec{k})}}.
\end{equation}

Secondly, let us fix a wave vector $\vec{k}_0$ and an arbitrary real number $\lambda^{(\vec{k}_0)}_0$ and consider the state $\psi \in {\cal H}_{p-poly}$ given by 
\begin{equation}
\psi= \prod_{\vec{k}} \tilde{\Psi}^{(\vec{k})} , \qquad \tilde{\Psi}^{(\vec{k})} = \left\{ \begin{array}{c} N_{\lambda^{(\vec{k}_0)}_0}, \quad \mbox{If} \quad \vec{k}=\vec{k}_0 , \\ 1, \qquad \mbox{If} \quad \vec{k}\neq \vec{k}_0 , \end{array}\right.  \label{teststate}
\end{equation}
\noindent where particular emphasis should be made on the discontinuity of the state $ \tilde{\Psi}^{(\vec{k})}$ as a formal function of the variable $\vec{k}$. Recall now that the cyclic state derived from the GNS-construction (usually considered as the vacuum) which we denote as $\psi_0$ is given by
\begin{equation}
\psi_0 = \prod_{\vec{k}} \tilde{\Psi}^{(\vec{k})}_0 , \qquad \tilde{\Psi}^{(\vec{k})}_0 = 1,  \qquad \forall \quad \vec{k} \in \mathbb{R}^3.
\end{equation}

 An element of the Weyl algebra ${\cal W}$ acts on the cyclic state $\psi_0$ as 
\begin{equation}
\pi[\widehat{W}(0,g_{\pi})] \psi_0 = \pi[\widehat{W}(0 , \{ g_{\vec{k}} \} )] \psi_0 = \pi[\widehat{W}(0 , \{ g_{\vec{k}} \} )] \prod_{\vec{k}} \tilde{\Psi}^{(\vec{k})}_0 = \prod_{\vec{k}} \pi[\widehat{W}(0 , g_{\vec{k}} )]  \tilde{\Psi}^{(\vec{k})}_0 = \prod_{\vec{k}} N_{g_{\vec{k}}}. 
\end{equation}
Due to the $g_{\vec{k}}$ are the Fourier transforms of Schwartz functions they are also Schwartz functions on their argument $\vec{k}$. In the case of states given by (\ref{teststate}) this is not longer valid. As result, the state $\psi$ cannot be obtained with the action of the Weyl algebra on $\psi_0$. In others words, the vacuum is not a cyclic vector on ${\cal H}_{poly}$ and therefore, the representation is not irreducible. It is the arbitrariness of the values of $\lambda^{(\vec{k})}$ in the Hilbert space ${\cal H}_{poly}$ the step that leads to the reducibility of the representation. We can add that this intuitive and often used derivation of the Hilbert space, lacks of mathematical rigor and renders difficult its implementation on curved space-times. As an attempt to improve it, we will offer in the next section, another way to derive the Hilbert spaces associated to the previous positive linear functionals $\omega_{\pi -pol}$. 

Let us analyze now the symmetry group associated to the PF representation. Recall that a diffeomorphism $f$ induces a transformation of the phase space points given by (\ref{FieldDif}-\ref{MomenDif}) and denoted by $T_f$. This transformation in turn, induces a transformation $A_{T_f}$ on the space ${\cal F}$ given in (\ref{TSObs}). Let us use this transformation $A_{T_f}$ in order to define an  automorphism $\alpha_f$ in the Weyl algebra ${\cal W}$ of the form
\begin{equation}
\alpha_f : {\cal W} \rightarrow {\cal W}; W(\lambda) \mapsto \alpha_f(W(\lambda)) := W(T_f(\lambda)),
\end{equation}
\noindent which explicitly takes the form
\begin{eqnarray}
W(\varphi_0; \pi_0) \mapsto W(\varphi'; \pi')&:=& W\left( \widehat{A} \varphi_0 + \widehat{B}\pi_0 ; \widehat{D} \varphi_0 + \widehat{E}\pi_0  \right). \label{AutoM}
\end{eqnarray}
 
 On the other hand, a group ${G}$ is called a symmetry group on a given algebra if there is a realization of ${G}$ by automorphisms $g \in { G} \mapsto \alpha_g \in \mbox{Aut} ({\cal W})$ such that
 \begin{equation}
 \omega(\alpha_g ({\cal W})) =  \omega({\cal W}), \label{SGDef}
 \end{equation}
\noindent where ${\cal W}$ correspond in this case to the Weyl $C^*$ algebra. This is connected with a unitary realization of the classical symmetry into the quantum description of the field \cite{Bratelli, Haag}. Following the definition of symmetry group, let us insert relation (\ref{AutoM}) in the singular states (\ref{CoordPol}) and (\ref{MomentumPol}). The $\varphi-$polarization given in (\ref{CoordPol}) yields
 \begin{eqnarray}
\omega_{\varphi-pol}(\widehat{W}(\varphi'_0 ;\pi'_0) )&=& \delta_{\varphi'_0, 0 } =  \delta_{\tilde{A} \varphi_0 + \tilde{B} \pi_0 , 0 } .
\end{eqnarray}

We observe that the condition (\ref{SGDef}) implies that those diffeomorphisms such that the relation
 \begin{eqnarray}
\delta_{\varphi_0, 0} = \delta_{\varphi'_0, 0}, \label{CSGFRep}
\end{eqnarray}
\noindent holds are elements of the symmetry group of the representation. Relation (\ref{CSGFRep}) implies that the diffeomorphisms acting on initial data $\varphi$ moves the data on the same hypersurface. Recall that
\begin{equation}
\varphi'(0; \vec{x}') = \cos\left[ t(0;\vec{x}') \sqrt{\widehat{\Theta}(\vec{x}(0;\vec{x}'))}\right] \varphi(\vec{x}(0;\vec{x}')) + \widehat{\Theta}(\vec{x}(0;\vec{x}'))^{- \frac{1}{2}} \sin\left[ t(0;\vec{x}') \sqrt{\widehat{\Theta}(\vec{x}(0;\vec{x}'))} \right] \pi(\vec{x}(0;\vec{x}')).
\end{equation}
\noindent Only diffeomorphisms given by $t' = t$ and $\vec{x}' = \vec{x}'(\vec{x})$ satisfy the condition (\ref{CSGFRep}) for any initial Cauchy hypersurface. We will call a diffeomorphism fulfilling this condition a `tangential' diffeomorphism and we will denoted by $\mbox{Diff}^{T}(\mathbb{R}^4)$ the group of all such diffeomorphisms.

Let us consider now the symmetry group of the $\omega_{\pi-pol}$ singular algebraic state given in (\ref{MomentumPol}). The expression (\ref{SGDef}) within the $\pi$-polarization yields the condition
\begin{eqnarray}
\pi_0 &=& \widehat{D} \varphi_0 + \widehat{E}\pi_0
\end{eqnarray}
Similarly to the previous polarization, the state $\omega_{\pi-pol}$ is invariant only when $\widehat{D} = 0$ and $\widehat{E} = 1$ which leads to the diffeomorphisms of the form 
\begin{equation}
\mbox{SDiff}(\mathbb{R}^4) = \{ t' = t, \; \vec{x}' = \vec{x}'(\vec{x}), \quad \mbox{such that} \quad \sqrt{\det{q'}} =1 \},
\end{equation} 
\noindent which are diffeomorphisms that induces unit Jacobian \cite{Khesin}.

$\mbox{SDiff}(\mathbb{R}^4)$ group contains the Euclidean group ${\cal E}$ as a subgroup. This is the symmetry group which left invariant the PF representation given in \cite{Viqar}. We know that the Euclidean group is a subgroup of the Poincar\'e group  $\mbox{Diff}(\mathbb{R}^4) \supset {\cal P} \supset {\cal E}$ and as we already mentioned, it is also contained in $\mbox{SDiff}(\mathbb{R}^4)$. Therefore, it is the intersection between both groups ${\cal P}\, \cap \, \mbox{SDiff}(\mathbb{R}^4) = {\cal E}$. This implies that there are classical symmetries (for instance boosts and time-translations) which are not implemented as unitary operators in the PFR of the real scalar field. This is a pathological aspect of the PFR which we hope might be circumvented when the dynamical description is invoked. Additionally, notice that tangent deformations of the hypersurface will be represented as unitary operators in a clear contrast to the standard Fock representation. As a result, we can conclude that PFR is a singular representation of the CCR of the scalar field which is not unitary equivalent to the standard Fock quantization based on $J^{(P)}$. 


\section{Relation with the Fock representation} \label{Comparisson}

The goal of this section is to compare PFR with the standard Fock representation. In order to do so, we derive the Polymer Hilbert space of the PFR within the GNS-construction. That is to say, the Hilbert space in which the vacuum is invariant under the group SDiff($\mathbb{R}^4$) of the momentum polarization of the field.

To begin with, consider the standard Weyl algebra ${\cal W}$ defined in the previous section. Recall that its generators are labeled with elements of the symplectic space $(\Gamma, \Omega)$.  From now on we change the notation to that used in the literature and in order to avoid confusion with our previous notation. The elements of the Weyl algebra of observables, corresponding to the Standard representation and PFR will be denoted as $W(g,f)$ where $g \in {\cal S}^{(0)}(\mathbb{R}^3)$ and $f \in {\cal S}^{(1)}(\mathbb{R}^3)$. From now on $f$ is no longer a diffeomorphism.

  Recall that PFR is given in the momentum polarization as derived from the state (\ref{MomentumPol}) or (\ref{ASPiPol}) and therefore, instead of considering the algebra ${\cal U}$ employed in the standard representation \cite{VelhinhoMTP} and which gives rise to the field configuration, let us consider the abelian algebra ${\cal V}$ given as follows.

Consider the set $V$ consisting of a finite number of functions $V= \{ g_j \}^N_{j=1}$ where $g_j \in {\cal S}^{(0)}(\mathbb{R}^3)$. For a given set $V$, the vector space formed with arbitrary finite linear combination of elements
\begin{equation}
\widehat{V}_{g}:= \widehat{W}(g,0), \label{HolM}
\end{equation}
\noindent can be endowed with an $\star$ abelian unital algebra structure with the multiplication taking the form
\begin{equation}
\widehat{V}_{g_1} \widehat{V}_{g_2} = \widehat{V}_{g_1 + g_2}, \qquad \widehat{V}^\dagger_{g} = \widehat{V}_{-g}.
\end{equation}
\noindent Let us endow this algebra with the supremum norm and complete it with this norm. The resulting algebra, denoted by ${\cal V}$, is indeed a sub-algebra of the Weyl algebra ${\cal W}$ used in the standard quantization. Classically, the generators $\widehat{V}_g$ can be seen as the functionals $V_g[\pi] = e^{-i \int d^3\vec{x} \, g \, \pi}$. 

In the standard Fock representation, Poincar\'e invariant state (\ref{Fstate}) is used to obtain the Hilbert space given by ${\cal H}_{Fock} = L^2({\cal S}^{'(1)}(\mathbb{R}^3), d\mu[\pi'])$ as we derive in the appendix. A particular step in the derivation of ${\cal H}_{Fock}$ is the use of the Bochner-Minlos theorem. This theorem allow us to derive the quantum configuration space as ${\cal S}^{'(1)}(\mathbb{R}^3)$ and the measure $d\mu[\pi']$ which is a Gaussian regular measure. However, the regularity (among other features) of the algebraic state (\ref{Fstate}) is in the core of the derivation. In the present case of the state given in (\ref{ASPiPol}), the regularity is absence and therefore, we are induced to applied Gel'fand spectral theory to construct the Hilbert space associated to the momentum polarization. We will follow similar  steps to those given in \cite{AshtekarLewansdoski} as well as \cite{LQProgram3} used in the Polymer Loop representation.

 First, consider the space $\Delta_{{\cal V}}$ which is the Gel'fand spectrum of ${\cal V}$, i.e., the set of all non-zero linear homomorphism 
\begin{equation}
\chi: {\cal V} \rightarrow \mathbb{C}; v \mapsto \chi[v].
\end{equation}
\noindent The spectrum $\Delta_{{\cal V}}$ of the abelian $C^*$ algebra ${\cal V}$ is a compact topological Hausdorff space with respect to some regular Borel measure $\mu_{PF}$ \cite{Thiemann1} which will be specified further below. Notice as usual that ${\cal S}^{(1)}(\mathbb{R}^3) \subset \Delta_{\cal V}$, i.e., the classical (momentum) space is densely embedded in the quantum (momentum) space \cite{Thiemann1}.  
 
 Let us consider now the Gel'fand isomorphism given by
 \begin{equation}
 \breve{(\cdot)}:  {\cal V} \rightarrow C(\Delta_{{\cal V}}); v \mapsto \breve{v}, \quad \mbox{such that} \quad \breve{v}[\chi]:= \chi[v], \quad \mbox{and} \quad \chi \in \Delta_{{\cal V}}. 
 \end{equation}
 The map $\breve{(\cdot)}$ relates the algebra ${\cal V}$ with that of the $C^*$ abelian algebra of continuous functions on $\Delta_{{\cal V}}$, denoted by $C(\Delta_{{\cal V}})$ \cite{LQProgram3}. Due to $\Delta_{\cal V}$ is a compact Hausdorff space we can define the regular measure $\mu_{PF}$ 
\begin{equation}
\omega_{PF}(v) = \int_{\Delta_{\cal V}} d\mu_{PF} \, \breve{v}[\chi],
\end{equation}
\noindent particularly,
\begin{equation}
\omega_{PF}(V_g)=\delta_{g,0} = \int_{\Delta_{\cal V}} d\mu_{PF} \, V_g,
\end{equation}
\noindent and the Hilbert space is of the form
\begin{equation}
{\cal H}_{PF}=L^2(\Delta_{\cal V}, d\mu_{PF} ).
\end{equation}

Although we are following the steps given in \cite{Thiemann1}, we are ignoring the proofs that might be required on this derivation. For instance, we will not pay attention to the specific nature of the spectrum $\Delta_{\cal V}$. Instead of that, we will follow an heuristic derivation and mention that it is  probably related to the Stone-\v{C}ech compactification of ${\cal V}$, named $\overline{\cal V}$.  In this spirit, we can consider that there is a map between $\overline{\cal V}$ and $\Delta_{\cal V}$ which let us consider general elements of the spectrum as the functional 
\begin{equation}
\chi(v) = \sum_j v_j \, \chi(V(g_j)) = \sum_j v_j \; e^{-i \int_{\mathbb{R}^3} g_j \pi }. \label{PSpectForm}
\end{equation}

\noindent In this way, our description resembles the techniques used in LQG formalism and on the toy models, particularly, polymer quantum mechanics in momentum polarization.

With these considerations, the Gel'fand isomorphism renders the elements on $C(\Delta_{\cal V})$ to be of the form
\begin{equation}
\breve{v}[\chi] =\chi(v) = \sum_j v_j \, \chi(V(g_j)) = \sum_j v_j \; e^{-i \int_{\mathbb{R}^3} g_j \pi } = : \Psi[\pi], \label{HSElement}
\end{equation}
\noindent and the similarity with the mechanical system in momentum polarization becomes more evident. Additionally, the generators of the algebra ${\cal V}$ can be represented as
\begin{equation}
(\widehat{V}_{g} \breve{v})[\chi]:= \chi(V_{g}) \breve{v}[\chi] = e^{-i \int_{\mathbb{R}^3} g \pi } \breve{v}[\chi],
\end{equation}

\noindent and, again, in a similar manner to the mechanical systems, the momentum field operator can not be recovered
\begin{equation}
\omega_{PF}(V_g) = \int_{\Delta_{\cal V}} d\mu_{PF} \, V_g = \int_{\Delta_{\cal V}} d\mu_{PF} \, \overline{\breve{v}}_0[\chi] \, \widehat{V_g} \, \breve{v}_0[\chi] = \langle 0 | \widehat{V_g} | 0 \rangle_{PF} =\delta_{g,0},
\end{equation}
\noindent where $\breve{v}_0[\chi]$ is the cyclic vector which takes the form $\breve{v}_0[\chi] =1$. As can be notice, this Hilbert space ${\cal H}_{PF}$ is not the same as (\ref{HPPoly}) which is the reported in \cite{Viqar}. 

The field operator can be represented using the basis element $V_g(\pi) := e^{-i \int_{\mathbb{R}^3} g \pi }$ of the Hilbert space ${\cal H}_{PF}$ as
\begin{equation}
\widehat{U}_f  \; e^{-i \int_{\mathbb{R}^3} g \pi } =  e^{-i \int_{\mathbb{R}^3} g ( \pi - f) },
\end{equation}
\noindent and as a result, the algebra multiplication is closed
\begin{equation}
\left(\widehat{U}_f \, \widehat{V}_g\right) [e^{-i \int_{\mathbb{R}^3} g' \pi }] = \left(e^{i \int_{\mathbb{R}^3} f g} \; \widehat{V}_g \, \widehat{U}_f \right) [e^{-i \int_{\mathbb{R}^3} g' \pi }] .
\end{equation}

 Let us now attend the question of how do Fock space  of the standard quantization of the scalar field arise from this non-Fock PFR of the same field? The entire answer to this question must await  the construction of the full physical Hilbert space after considering the dynamic of the quantum states. However, the procedure employed in this section, resembles that of the $U(1)$ holonomy algebras given in \cite{Varadarajan2, Varadarajan3} and pursue illuminate some facets of the relation between PFR and the Poincar\'e invariant Fock representation.
 
 Consider the amplitude of the operator $\widehat{V}(g)$ using the vacuum vector $\Psi_0$ in the standard representation $\langle \Psi_0 | \widehat{V}(g) \Psi_0 \rangle$ given by
\begin{equation}
\langle \Psi_0 | \widehat{V}(g) \Psi_0 \rangle = \int_{{\cal S}^{'(1)}} d\mu_G \, \overline{\Psi}_0 \, (\widehat{V}(g) \Psi_0) = \omega_{J^{(P)}}(\widehat{V}(g)) = e^{-\frac{1}{4} \int_{t_0} d^3\vec{x} g (-\Delta + m^2)^{1/2} g}. \label{MCond}
\end{equation}

If we consider the field operator, this amplitude value can be obtained for the vacuum state $\Psi_0$ as
\begin{equation}
 \omega^2_{J^{(P)}}(\widehat{V}(g)) \langle \Psi_0 |e^{i \int_{t_0} \widehat{\varphi} [-i (-\Delta + m^2)^{1/2} g]} | \Psi_0 \rangle  = e^{-\frac{1}{4} \int_{t_0} d^3\vec{x} g (-\Delta + m^2)^{1/2} g}. \label{FCond}
\end{equation}

Combining both relations (\ref{MCond}) and (\ref{FCond}), we obtain that the vacuum in the standard representation, is the only state satisfying the condition
\begin{equation}
\widehat{P} \, \Psi_0 = 0, \qquad \widehat{P} := \widehat{V}(g) -  \omega^2_{J^{(P)}}(\widehat{V}(g)) e^{i \int_{t_0} \widehat{\varphi} [-i (-\Delta + m^2)^{1/2} g]}, \label{PCondition}
\end{equation}
\noindent for every $g \in {\cal S}^{(0)}(\mathbb{R}^3)$. This condition can be seen as an exponential version of the Poincar\'e invariance condition of the vacuum but in terms of the ${\cal V}$ generators \cite{Varadarajan3}. 

In the case of the $U(1)$ and scalar field quantized within Loop Quantization techniques, the equivalent $\widehat{P}$ operator is defined using an isomorphism between the Loop Hilbert space and an auxiliar Hilbert space constructed by smearing the holonomies of the gauge fields as well as the point holonomies of the scalar field \cite{AshtekarLewansdoski,Varadarajan2, Varadarajan3, AbhayLew}. In the present case, $\widehat{P}$ is naturally defined on ${\cal H}_{PF}$ due to both, standard and PFR use the same algebra of observables but with a different positive linear functional. In other words, condition (\ref{PCondition}) can be implemented on the PFR but is not satisfied for any (non-trivial) element of $\Omega \in {\cal H}_{PF}$ 
\begin{eqnarray}
 \widehat{P} \, \Omega &=&  \left( \widehat{V}(g) -  \omega^2_{J^{(P)}}(\widehat{V}(g)) e^{i \int_{t_0} \widehat{\varphi} [-i (-\Delta + m^2)^{1/2} g]} \right)  \Omega , \nonumber \\
 \widehat{P} V_{g'}(\pi)&=& V_{g + g'}(\pi) -  \omega^2_{J^{(P)}}(\widehat{V}(g))  e^{\int g' (- \Delta + m^2 )^{1/2} g} V_{g'}(\pi) \neq 0.
\end{eqnarray}
Essentially, the states $V_{g+g'}$ and $V_g$ are linearly independent orthogonal states. In the case of Loop Quantized theories \cite{AshtekarLewansdoski,Varadarajan2, Varadarajan3, AbhayLew}, this result suggest to look for states on the dual ${\cal D}^*$ of the space ${\cal D}$. The space ${\cal D}$ is the (dense) set of finite linear of charged network states. A similar dense space is available in the PFR as can be seen as follows. 
 
Recall that in the GNS-construction, the representation of the algebra ${\cal V}$ as operators over ${\cal H}_{PF}$ acting on the cyclic vector $\breve{v}_0[\chi]$ yields a dense space ${\cal D} \subset {\cal H}_{PF}$ which is formed by the finite linear combinations of the form
\begin{equation}
{\cal D} := \left\{ \sum_{g_j} c_{g_j} V_{g_j}(\pi) \right\}.
\end{equation}
Its (algebraic) dual ${\cal D}^*$, with elements $\Phi[\Omega] \in {\cal D}^*$, is given by the complex linear maps on ${\cal D}$. Let us define the dual basis elements $\Phi_g$ such that $\Phi_g[V_{g'}] = \delta_{g,g'}$. Every element of the dual can be given as a formal sum 
\begin{equation}
\Phi := \sum_g c_g \Phi_g, \qquad \Phi[V_{g'}] = \sum_g c_g \delta_{g,g'} = c_{g'}. \label{DualElement}
\end{equation}

Consider the arbitrary element $\Phi$ of the dual ${\cal D}^*$ given in (\ref{DualElement}) and let us insert the Poincar\'e condition (\ref{PCondition}) yielding
\begin{eqnarray}
\Phi[\widehat{P} V_{g'}] &=& \sum_{\tilde{g}} c_{\tilde{g}} \Phi_{\tilde{g}}[\widehat{P} V_{g'}] = \sum_{\tilde{g}} c_{\tilde{g}} \Phi_{\tilde{g}}\left[ V_{g + g'} -  \omega^2_{J^{(P)}}(\widehat{V}(g))  e^{\int g' (- \Delta + m^2 )^{1/2} g} V_{g'} \right] , \nonumber \\
&=& \sum_{\tilde{g}} c_{\tilde{g}} \Phi_{\tilde{g}}[ V_{g + g'}] -  \omega^2_{J^{(P)}}(\widehat{V}(g))  e^{\int g' (- \Delta + m^2 )^{1/2} g} \sum_{\tilde{g}} c_{\tilde{g}} \Phi_{\tilde{g}}[ V_{g'}] , \nonumber \\
&=&  c_{g + g'} -  \omega^2_{J^{(P)}}(\widehat{V}(g))  e^{\int g' (- \Delta + m^2 )^{1/2} g} c_{g'}.
\end{eqnarray} 

We can now look for states $\Phi$ on the dual space ${\cal D}^*$ such that $\Phi[\widehat{P} V_{g'}] =0$, i.e., solve the equation
\begin{equation}
c_{g + g'} -  \left( e^{-\frac{1}{4} \int  g (-\Delta + m^2)^{1/2} g} \right)^2  e^{\int g' (- \Delta + m^2 )^{1/2} g} c_{g'} =0,
\end{equation}
\noindent for the coefficients $c_g$. Notice that since this equation is linear and homogeneous, the solution will be ambiguous by an overall constant but this ambiguity can be fixed by setting the coefficient $c_0$ to be unity. Then the solution takes the form
\begin{equation}
c_g = e^{-\frac{1}{2} \int g (-\Delta + m^2)^{1/2} g}.
\end{equation}

It may be verified that the dual state with this coefficients is the unique (up to an overall constant) solution to the condition $\Phi_0[\widehat{P} V_{g'}] =0$ of the form
\begin{equation}
\Phi_0 = \sum_g e^{-\frac{1}{2} \int g (-\Delta + m^2)^{1/2} g} \Phi_g,
\end{equation}
\noindent where the sum must be understood as a formal infinite sum over the (now) {\em continuous variable} $g$ in order to obtain the same result $\Phi_0[\widehat{P} v] =0$ for any element of $v \in {\cal D}$. The Gel'fand triplet reads as 
\begin{equation}
{\cal D} \subset {\cal H}_{PF} \subset {\cal D}^*  \ni \Phi_0,
\end{equation}
\noindent and the algebra ${\cal V}$ and ${\cal U}$ can be extended to the dual as follows
\begin{eqnarray}
(\widehat{V}_g \Phi)[v] &=& \Phi[\widehat{V}^\dagger_g \, v] = \Phi[\widehat{V}_{-g} \, v] , \\
(\widehat{U}_f \Phi)[v] &=& \Phi[\widehat{U}^\dagger_f \, v] = \Phi[\widehat{U}_{-f} \, v] .
\end{eqnarray}

The action of the ${\cal V}$ algebra on the dual state $\Phi_0$ provides a dense subspace $ {\cal L}^*$ of the dual space ${\cal D}^*$. Any element of ${\cal L}^*$ is of the form $\sum_{g_j} c_{g_j} \widehat{V}_{g_j} \Phi_0$ for some complex number $c_{g_j}$ and the summation running with a finite number of terms. Typically, ${\cal D}^*$ and hence ${\cal L}^*$ are not equipped with an inner product. Again, following \cite{AshtekarLewansdoski,Varadarajan2, Varadarajan3, AbhayLew}, an inner product, implementing the classical reality conditions on the quantum operators can be given as
\begin{equation}
 \langle \widehat{W}(g,f) \Phi_0[v] |  \widehat{W}(g',f') \Phi_0[v] \rangle_{P} :=  \omega_{J^{P}}(\widehat{W}(g',f') \widehat{W}(-g,-f)). \label{IProd}
 \end{equation}
The Cauchy completion of ${\cal L}^*$ with respect to this inner product, results in a Hilbert space $({\cal L}^* , \langle \cdot \rangle_{P})$ unitarily equivalent to the Hilbert space of the standard quantization ${\cal H}_{Fock} = L^2({\cal S}^{'(1)}(\mathbb{R}^3), d\mu)$. The GNS-construction guarantees a unitary representation of the elements of the symmetry group $\mbox{SDiff}(\mathbb{R}^4)$ on ${\cal H}_{PF}$. Each of this operators can be extended to the dual ${\cal D}^*$ and of course, to ${\cal L}^*$. However, the inner product (\ref{IProd})  ensures that only those element of $\mbox{SDiff}(\mathbb{R}^4) \cap {\cal P}$ act as unitary operators. Conversely, the elements of the Poincar\'e group  related to a time transformation (boosts and time translations) are representation as non-unitary operators on ${\cal H}_{PF}$. These operators admit an extension to the Hilbert space $({\cal L}^*,  \langle , \rangle_{P})$ which results to be unitary as is expected.

\section{Discussion}  \label{discusion}

Polymer quantization of the Fourier modes of the real scalar field is a rather different quantization of this field. It is a hybrid quantization that uses the standard Weyl algebra of observables and an algebraic state which mimics that of Loop quantization of the same field. This mixture serves as an arena to understand some aspects of the standard Fock quantization in regard to the Loop quantization, particularly, the relation with symmetry groups such that the Poincar\'e and the diffeomorphism groups.

The construction of the polymer Fourier representation given in \cite{Viqar} lacks of the analysis of its own symmetry group and a more precise relation with the Poincar\'e group unitarily implemented in the standard quantization. This work attends these questions at a kinematical level and from, an algebraic description.

First, we derived the algebra of observables ${\cal W}$(Weyl algebra) of the standard quantization. We followed the derivations given in \cite{JeronimoCQ1, JeronimoCQ2, Wald, VelhinhoMTP}, particularly, we used the Poincar\'e-Fock invariant complex structure of the standard quantization to express the algebraic state of the field theory (\ref{Fstate}) in terms of the algebraic states of each Fourier modes (\ref{FstateMod}). Additionally, we implemented an action of the diffeomorphism group on the phase space $\Gamma$ as well as in the symplectic space ${\cal F}$.

Secondly, we used some of the results given in \cite{CorichiZapata} and we make the following proposal: {\em the replacement of the standard quantum harmonic oscillators (in momentum polarization) of each Fourier mode of the scalar field by its polymer analog, is equivalent to replace the Poincar\'e invariant algebraic state (\ref{Fstate}) by a new algebraic state given by (\ref{MomentumPol})}. This declaration, paves the way to the algebraic construction of the Polymer Fourier Representation and allow us to study the group symmetries of such algebraic state. This proposal is inspired by the fact that as was showed in \cite{CorichiZapata}, the polymer quantization of the harmonic oscillator can be seen as a singular limit of the parameter of the complex structure. The net result is a replacement of the standard algebraic state (\ref{MechState}) by a non-regular algebraic state (\ref{SingMechState}).

There are two immediate consequences of this proposal in the analysis of this singular algebraic state (\ref{MomentumPol}). The first is that the polymer Hilbert space given in \cite{Viqar} yields a reducible representation of the Weyl algebra. A different polymer Hilbert is required in order to fix a irreducible space for the observables and, naturally,  due to the standard Poincar\'e invariant representation of the scalar field is also irreducible, the appropriate scenario for the comparison of both representations is given when both are irreducible representations of the same Weyl algebra. The second result is that the group which left invariant the polymer algebraic state (\ref{MomentumPol}) is the sub-group $\mbox{SDiff}(\mathbb{R}^4)$ of spatial diffeomorphisms whose elements preserve the spatial volume. 

This new algebraic state unveils a new representation of the Weyl algebra of observable of the real scalar field, named PFR, which is a singular representation of the CCR. In the present case of the algebraic state (\ref{MomentumPol}), the momentum operator of the field is not well defined and therefore no number particle operator can be obtained. This is a typical feature of Loop quantization and is a result of the non-regularity of the algebraic state.

The vacuum state of the PFR is not Poincar\'e invariant, specifically, not invariant under a boost or a time translation. The other elements of the Poincar\'e group can be represented as unitary operators on the polymer Hilbert space ${\cal H}_{PF}$. Even more, any diffeomorphism affecting the time parameter (i.e., $t' \neq t$) cannot be promoted to a unitary operator on ${\cal H}_{PF}$. Two observers related by this kind of diffeomorphism will measure different amplitudes of the vacuum state. An example of this type of diffeomorphism is the Rindler transformation which gives rise to the Unruh effect \cite{Rindler}. The Rindler observer is not an inertial observer and consequently, an observer within the standard Fock quantization measures a different value of the vacuum amplitude than the one measured by a Rindler observer in the same representation \cite{Halvorson}.  The same result will be obtained in the PFR. A Rindler observer within the PFR will measure a different  vacuum amplitude to that measure by an inertial observer also in the PFR.

However, it is worth noticing that this conclusion is entirely at a kinematical level. When the dynamics is invoked in this PFR, the usual techniques \cite{Jackiw} must be mixed with those employed in \cite{AshtekarWillis} for the mechanical harmonic oscillator. It might happen that the resulting dispersion relation turns out to be different to that obtained in \cite{Viqar}. For example, we can consider that instead of using a global(does not depends on the spacetime points), constant(does not depends on the Fourier modes) and fixed parameter $M_{\star}$ for the polymer Hamiltonian, a different parameter $M_{\star}(\vec{x}) \in {\cal S}^{(0)}(\mathbb{R}^3)$ can be considered. In doing so, the combination of holonomies $\widehat{V}_g$ used for the replacement of the term $\widehat{\pi}^2(\vec{x})$ in the Hamiltonian density, will now be well defined according to the Hilbert space ${\cal H}_{PF}$ derived in this work. Although at this level our results are consistent with those reported in \cite{VLouko, Kajuri,GolamSardar}, it would be interesting to study if such a dynamical description yields the same results as in \cite{VLouko, Kajuri, GolamSardar}.

Finally, in the third place, we compared in the last Section both, the Poincar\'e invariant representation and the irreducible PFR. The Hilbert space ${\cal H}_{PF}$ was formally introduced together with the representation of the algebra of observables ${\cal U}$ and ${\cal V}$. In this case, the Gel'fand spectral theory was employed. We avoid the characterization of the spectrum $\Delta_{\cal V}$ and proposed to be isomorphic to that formed with functional given by (\ref{PSpectForm}). This assumption was made based on the similitude between our derivation and that of the polymer harmonic oscillator. 

A noteworthy feature used in the comparison is that both representations are based on the same Weyl algebra ${\cal W}$ and what is different are the algebraic states. Comparisons of the loop representation of the $U(1)$ gauge field \cite{Varadarajan2, Varadarajan3, AbhayLew} and the loop representation of the scalar field \cite{AshtekarLewansdoski} with their standard Fock representation uses an auxiliary Hilbert space ${\cal H}_{(r)}$. This Hilbert space is introduced because the holonomies (or point holonomies for the scalar field) are not well defined in the standard Fock quantization. In the PFR this is not the case and this auxiliary Hilbert space is not required.  A direct consequence is that the Poincar\'e condition, encoded in the operator $\widehat{P}$ can be naturally defined on the Hilbert space ${\cal H}_{PF}$. Of course, there is not Poincar\'e invariant vector on ${\cal H}_{PF}$ and therefore, we are forced to look into the dual space ${\cal D}^*$. Remarkably, the dual space ${\cal D}^*$ admits a Poincar\'e invariant vector which is later used to define the Fock Hilbert space $({\cal L}^*, \langle, \rangle_P)$. 

In this construction, the elements of the symmetry group $\mbox{SDiff}(\mathbb{R}^4)$ of the PFR are unitary operators acting on ${\cal H}_{PF}$. On the other hand, the elements of the Poincar\'e group which are not in the group $\mbox{SDiff}(\mathbb{R}^4)$, i.e., boost and time translations, are not unitary operators on this Hilbert space. However, every operator acting on ${\cal H}_{PF}$ can be promoted to an operator acting on the dual space ${\cal D}^*$. Recall that ${\cal L}^* \subset {\cal D}^*$ hence, there is an action of the operators of ${\cal H}_{PF}$ acting on ${\cal L}^*$. 

The space ${\cal L}^*$ is not an inner product space and therefore, the unitary property of operators coming from ${\cal H}_{PF}$ is not well defined. Nevertheless, this space can be endowed with the inner product (\ref{IProd}). In this inner product, a new notion of unitarity for the operators is implemented. As a result, elements of $\mbox{SDiff}(\mathbb{R}^4) \backslash {\cal P}$ are no longer unitary operators and the resulting symmetry group is the Poincar\'e group.

As a final remark, let us mention that there are, of course, some aspects to be taken with a pinch of salt. In our opinion, the first is our proposal of the polymer algebraic state mimicking the replacement of the quantum harmonic oscillators by its polymer version and the second consists on the nature of the spectrum $\Delta_{\cal V}$ which allow us to write an element of the Hilbert space ${\cal H}_{PF}$ in the form (\ref{HSElement}). The first might require a bit more discussion and the second, more mathematical rigour could be invoked, probably in future works.

\section*{ACKNOWLEDGMENTS}
We are grateful to Alejandro Corichi, Juan Reyes and Hugo Morales-T\'ecotl for important comments along the preparation of this work. Angel Garcia-Chung acknowledges the total support from DGAPA-UNAM fellowship. The authors acknowledge partial support from CONACYT project 237503 and DGAPA-UNAM grant IN 103716.

\section*{Appendix}

In this section we will derive the momentum polarization of the real scalar field in the standard Poincar\'e invariant quantization. We will follow the steps given in \cite{JeronimoCQ2}. 

Consider the functional space $\Psi[\pi']$, where $\pi' \in {\cal S}'^{(1)}(\mathbb{R}^3)$ is a distribution of weight one. Let the momentum operator be represented as a multiplicative operator
\begin{equation}
\widehat{\pi}[g] \Psi[\pi'] = \left( \int d^3 \vec{x} \, g \, \pi' \right) \Psi[\pi'].
\end{equation}

Consider now the positive linear functional associated to the Fock state $\omega_{J^{(P)}}$ applied to the generators of the algebra ${\cal V}$. The amplitude is given by
\begin{equation}
\omega_{J^{(P)}}\left( \widehat{W}(g,0) \right) = e^{- \frac{1}{4} \int d^3 \vec{x} g \, \widehat{\Theta}^{\frac{1}{2} } \, g },
\end{equation}
\noindent and can be written in terms of the inner product of the Hilbert space given by
\begin{equation}
e^{- \frac{1}{4} \int d^3 \vec{x} g \, \widehat{\Theta}^{\frac{1}{2} } \, g } = \int_{{\cal S}^{'(1)}(\mathbb{R}^3)} d\mu(\pi') \, \overline{\Psi}_0[\pi'] \, e^{-i \int d^3\vec{x} \, \pi' \,  g} \, \Psi_0[\pi'],
\end{equation}
\noindent where the cyclic vector $\Psi_0$ is of the form $\Psi_0[\pi'] =1$. Bochner-Minlos theorem implies that the measure is a Gaussian type measure as is well known in the field polarization, i.e., we can naively write
\begin{equation}
`` d\mu(\pi') = e^{- \int d^3 \vec{x} \pi' \, \widehat{\Theta}^{- \frac{1}{2}} \pi'  } {\cal D}{\pi'}. ''
\end{equation}
This measure endows the following representation for the field operator
\begin{equation}
\widehat{\varphi}[f] \Psi[\pi'] = i \int d^3\vec{x} \left( f \frac{\delta}{\delta \pi'} -  \pi' \widehat{\Theta}^{- \frac{1}{2}} f \right) \Psi[\pi']
\end{equation}
\noindent and can be verified that this representation satisfies the CCR.

The exponential version of the field operator (the Weyl element $\widehat{U}_f$) over the vacuum state is given by
\begin{equation}
e^{i \widehat{\varphi}[f]} \Psi_0[\pi '] = e^{- \frac{1}{2} \int f \, \widehat{\Theta}^{- \frac{1}{2}} \, f } e^{\int \pi' \widehat{\Theta}^{- \frac{1}{2}} f} \Psi_0[\pi'], \label{VUop}
\end{equation}
\noindent and the amplitude of the momentum holonomy takes the form
\begin{eqnarray}
\int_{{\cal S}^{'(1)}} d\mu_G \, \overline{\Psi}_0 \, (\widehat{V}(g) \Psi_0) = e^{-\frac{1}{4} \int_{t_0} d^3\vec{x} \, g \, \widehat{\Theta}^{\frac{1}{2}} \, g} = \omega^2_{J^{(P)}}(\widehat{V}(g)) \langle \Psi_0 |e^{ \int \widehat{\varphi} [ \widehat{\Theta}^{1/2} g]} | \Psi_0 \rangle . \label{VHol}
\end{eqnarray}

Combining these relations (\ref{VUop}) and (\ref{VHol}) we obtain the Poincar\'e condition of the standard quantization of the real scalar field within the momentum polarization $\widehat{P} \Psi_0 = 0$ where the operator $\widehat{P}$ is given by
\begin{equation}
\widehat{P} := \widehat{V}(g) -  \omega^2_{J^{(P)}}(\widehat{V}(g)) e^{i \int \widehat{\varphi} [-i\widehat{\Theta}^{1/2} g]} .
\end{equation}

\end{document}